\documentclass[aps,prc,showpacs,twocolumn,nofootinbib,preprintnumbers]{revtex4-1}

\usepackage{longtable}
\usepackage{graphicx}
\usepackage{dcolumn}
\usepackage{bm}
\usepackage{amsmath,amssymb}

\newcommand{\bra}[1]{\langle \, #1 \, |}
\newcommand{\ket}[1]{| \, #1 \, \rangle}
\newcommand{\kket}[1]{\, #1 \, \rangle}

\newcommand{\re}{\text{Re }}
\newcommand{\im}{\text{Im }}

\newcommand{\physdim}[1]{\hspace{1ex} (\mathrm{#1})}

\begin{document}

\preprint{YITP-15-76}


\title{Structure of near-threshold quasi-bound states}


\author{Yuki~Kamiya}
\email[]{yuki.kamiya@yukawa.kyoto-u.ac.jp}
\affiliation{Yukawa Institute for Theoretical Physics, Kyoto University, Kyoto 606-8502, Japan}
\author{Tetsuo~Hyodo}
\email[]{hyodo@yukawa.kyoto-u.ac.jp}
\affiliation{Yukawa Institute for Theoretical Physics, Kyoto University, Kyoto 606-8502, Japan}


\date{\today}

\begin{abstract}
We study the compositeness of near-threshold quasi-bound states in the framework of effective field theory. From the viewpoint of the low-energy universality, we revisit the model-independent relations between the structure of the bound state and the observables in the weak binding limit. The effective field theory enables us to generalize the weak-binding relation of the stable bound states to unstable quasi-bound states with decay modes. We present the interpretation of the complex values of the compositeness for the unstable states. Combining the model-independent relation and the threshold observables extracted from the experimental data, we show that $\Lambda(1405)$ is dominated by the $\bar{K}N$ molecular structure and that $a_{0}(980)$ is dominated by the non-$\bar{K}K$ component.
\end{abstract}




\maketitle

\section{Introduction}
The quest for the composite nature of particles is a major subject in physics. Historically, the identification of the fundamental constituents of elementary particles has been a central issue in particle physics~\cite{Weinberg:1962hj,PTP29.877,Eichten:1983hw,Perelstein:2005ka,Agashe:2014kda}. Aside from the high energy frontier, the exploration of the composite structure of the excitations in various systems is vastly discussed. In nuclear physics, clustering of nucleons leads to the molecule-like structure of nuclei~\cite{PTPS52.89}. In condensed matter physics, the structure of quasiparticles is studied in the polaron-molecule transition~\cite{Schmidt:2011zu} as well as the molecules associated with the Feshbach resonances in cold atoms~\cite{Kohler:2006zz}.

Recently, the composite nature of hadrons in quantum chromodynamics (QCD) has received renewed interest. Hadrons are the color singlet asymptotic states formed by quarks and gluons in the nonperturbative regime of QCD. Triggered by the observations of exotic hadrons~\cite{Swanson:2006st,Brambilla:2010cs,Belle:2011aa,Aaij:2015tga}, much attention is paid to the hadronic molecular states, which are the composite systems of hadrons bounded by the interhadron interactions. For instance, the $\Lambda(1405)$ resonance is considered to be the $\bar{K}N$ molecule~\cite{Dalitz:1967fp,Hyodo:2011ur,Hall:2014uca,Miyahara:2015bya}. It should however be noted that the hadronic molecules cannot be distinguished from the ordinary hadrons by quantum numbers. A method to characterize the composite structure of hadrons without relying upon specific models is desired.

A remarkable achievement in this direction is the weak binding relation for the stable bound state~\cite{Weinberg:1965zz}. It is shown that the compositeness $X$, the probability of finding the composite structure in the wavefunction of the bound state, can be related to the observables as
\begin{align}
	a_{0} &= R\left\{ \frac{2X}{1+X} + {\mathcal O}\left(\tfrac{R_{\mathrm{typ}}}{R}\right)\right\} , \label{eq:comp-rel-bound} \\
	r_{e} &= R\left\{ \frac{X-1}{X} + {\mathcal O}\left(\tfrac{R_{\mathrm{typ}}}{R}\right)\right\} , \label{eq:comp-rel-bound2}
\end{align}
where the radius $R=1/\sqrt{2\mu B}$ is determined by the binding energy $B$ with the reduced mass $\mu$. The relations assert that the scattering length $a_{0}$ and the effective range $r_{e}$ are determined by the radius $R$ and the compositeness $X$, if the radius $R$ is much larger than the typical length scale of the interaction $R_{\rm typ}$. Thus, the compositeness of the weakly bound state is reflected in the observable quantities in a model-independent manner. As an application, the deuteron is shown to be a loosely bound state of two nucleons, rather than the compact six-quark state~\cite{Weinberg:1965zz}. It should be emphasized that this conclusion is drawn by a few observables ($B$, $a_{0}$, and $r_{e}$), without explicitly knowing the wavefunction of the deuteron.

Although these relations are general, the applicability is limited to the stable bound states, while possible candidates for the hadronic molecular states are the unstable particles with finite decay width. For the application to hadronic molecules, it is indispensable to generalize the relation to unstable states. There have been several studies of the compositeness of the unstable particles~\cite{Baru:2003qq,Baru:2010ww,Hyodo:2011qc,Aceti:2012dd,Hyodo:2013iga,Chen:2013upa,Hyodo:2013nka,Sekihara:2014kya,Guo:2015daa}, but the direct generalization of the model-independent relations~\eqref{eq:comp-rel-bound} and \eqref{eq:comp-rel-bound2} has not been obtained.

In this work, we derive the weak-binding relation based on the low-energy effective field theory (EFT), which is a universal approach to extract the long-wavelength dynamics~\cite{Heisenberg:1935qt,Weinberg:1979kz,Gasser:1985gg,Weinberg:1990rz,Leutwyler:1993gf,Bodwin:1994jh,Braaten:2004rn,Braaten:2007nq,Watanabe:2014fva}. The EFT formulation provides much clear derivation of the weak binding relation in comparison with the original work~\cite{Weinberg:1965zz}. Moreover, the use of the EFT formulation is essential to generalize the relation to the unstable quasi-bound states. Carefully examining the interpretation of the complex-valued compositeness, we use the generalized relation to study the structure of the candidates of the hadronic molecular states.

\section{Model-independent relations for stable bound states}
We consider the single-channel $s$-wave scattering of two distinguishable particles with a shallow bound state. As long as the energy of the system is sufficiently small, the microscopic details of the fundamental interaction are irrelevant, and the system can be described by the effective nonrelativistic quantum field theory with local interactions~\cite{Braaten:2004rn,Braaten:2007nq}. Inspired by the EFT for the low-energy nuclear force~\cite{Kaplan:1996nv}, we utilize the Hamiltonian $H=H_{\mathrm{free}} + H_{\mathrm{int}}$:
\begin{align}
H_{\mathrm{free}} &=\int d\bm{r}
\biggl[\frac{1}{2 M} \mathbf{\nabla} \psi^\dagger \cdot\mathbf{\nabla} \psi +\frac{1}{2 m} \mathbf{\nabla} \phi^\dagger \cdot\mathbf{\nabla} \phi \nonumber \\
&\quad + \frac{1}{2 M_{0}} \mathbf{\nabla}  B_0^\dagger \cdot{\mathbf \nabla} B_0 +  \nu_0 B_0^\dagger B_0
\biggr] ,\\
H_{\mathrm{int}} &= \int d\bm{r}
\left[
g_{0} \left( B_0^\dagger \phi\psi + \psi^\dagger\phi^\dagger B_0 \right) + v_0 \psi^\dagger\phi^\dagger \phi\psi
\right] ,
\end{align}
with $\hbar=1$. We introduce the momentum scale cutoff $\Lambda$, below which the EFT description is accurate. The cutoff value should be related with the typical length scale of the fundamental interaction as $R_{\rm typ}\sim 1/\Lambda$. For a given scale $\Lambda$, the bare parameters $g_{0}$, $v_{0}$, and $\nu_{0}$ are related to the low-energy observables.\footnote{In fact, this EFT is renormalizable, i.e., there is a sensible limit $\Lambda\to \infty$ with the observables being fixed. Here we keep the cutoff finite, because the compositeness is in general a renormalization-dependent quantity. The scale dependence disappears in the weak-binding limit, as will be discussed later.}

Let us consider the scattering of the two-body $\psi\phi$ system. Because of the phase symmetry of the interaction Hamiltonian, the eigenstate of the two-body problem can be expressed by the linear combination of the eigenstates of the free Hamiltonian:
\begin{align}
   \ket{\Psi} 
   &=c\ket{B_{0}}+\int\frac{d\bm{p}}{(2\pi)^{3}} \chi(\bm{p})\ket{\bm{p}}
   \label{eq:twobody} ,
\end{align}
where the bare state and the scattering states are defined by $\ket{B_{0}}=\tilde{B}_{0}^{\dag}(\bm{0})/\sqrt{\mathcal{V}}\ket{0}$ and $\ket{\bm{p}}=\tilde{\psi}^{\dag}(\bm{p})\tilde{\phi}^{\dag}(-\bm{p})/\sqrt{\mathcal{V}}\ket{0}$ with the creation operators $\tilde{\psi}^{\dag}(\bm{p})$, the vacuum $\ket{0}$, and $\mathcal{V}=(2\pi)^{3}\delta^{3}(\bm{0})$. With Eq.~\eqref{eq:twobody}, the two-body problem can be exactly solved.

The eigenvector of the bound state $\ket{B}$ with the binding energy $B$ is obtained by solving the Schr\"odinger equation $H\ket{B}=-B\ket{B}$. The elementariness $Z$ and the compositeness $X$ of the bound state are defined as
\begin{align}
Z &\equiv |\bra{B_{0}}\kket{B}|^{2},\quad 
X \equiv \int \frac{d\bm{p}}{(2\pi)^{3}} |\bra{\bm{p}}\kket{B}|^{2} .
\label{eq:normalization}
\end{align}
It follows from the normalization of the bound state $\ket{B}$ and the completeness relation that~\cite{Hyodo:2013nka}
\begin{align}
Z+X =1,\quad Z,X \in [0,1] .
\label{eq:ZXbound}
\end{align}
This ensures the probabilistic interpretation; $Z$ ($X$) is the probability of finding the bare state $\ket{B_{0}}$ (the scattering state $\ket{\bm{p}}$) in the physical bound state $\ket{B}$. The forward scattering amplitude of the $\psi\phi$ system $f(E)$ is obtained by solving the Lippmann-Schwinger equation as
\begin{align}
f(E) &=-\frac{\mu}{2\pi} \frac{1}{\left[v(E)\right]^{-1} - G(E) } \label{eq:amp-bound},
\end{align}
with $\mu=Mm/(M+m)$ and
\begin{align}
v(E) &=v_0 + \frac{g_0^2}{E - \nu_0}, \\
G(E) &= \frac{1}{2\pi^2} 
\int_{0}^{\Lambda}dp \frac{p^{2}}{E-p^2/(2\mu)+i0^{+}}.
\end{align}
As shown in Ref.~\cite{Sekihara:2014kya}, the compositeness $X$ of this type of amplitude can be written as
\begin{align}
X & = \{1+ G^2(-B) v^\prime (-B)\left[ G^\prime (-B)\right]^{-1}\}^{-1} \label{eq:X-bound} ,
\end{align}
where $v^{\prime}(E)=dv(E)/dE$ and $G^{\prime}(E)=dG(E)/dE$.

We are now ready to derive the weak-binding relation~\eqref{eq:comp-rel-bound}. When the binding energy is small, the radius $R=1/\sqrt{2\mu B}$ becomes large. We expand the scattering length $a_0 = -f(E=0)$ in powers of $1/R$. With the help of Eq.~\eqref{eq:X-bound} and the bound state condition $v(-B)G(-B)=1$, the coefficient of the leading order term $\mathcal{O}(R)$ can be expressed solely by $X$~\cite{Sekihara:2014kya}. The correction to the leading order term depends on the cutoff scale $\Lambda\sim 1/R_{\rm typ}$. In other words, the correction terms are suppressed by $R_{\rm typ}/R$ compared with the leading order term. The sufficient condition for the $1/R$ expansion of $a_{0}$ is the validity of the effective range expansion at the bound state pole, from which Eq.~\eqref{eq:comp-rel-bound2} follows.

The correction term of $\mathcal{O}(R_{\rm typ}/R)$ represents the model-dependent contribution to $a_{0}$ and $r_{e}$ which reflects the short-range behavior of the interaction. At the same time, because $R_{\rm typ}\sim 1/\Lambda$, the correction term controls the renormalization scale dependence of $Z$ and $X$. If the binding energy is so small ($R$ is so large) that the correction term is neglected, the compositeness $X$ can be determined by the observables with no dependence on the scale $\Lambda$. In this limit, the compositeness $X$ is invariant under the field redefinitions and determined only by observables. Equations~\eqref{eq:comp-rel-bound} and \eqref{eq:comp-rel-bound2} are thus mentioned as ``model-independent'' relations. 

It is instructive to recall the scaling limit $R_{\rm typ}\to 0$ with the scattering length being kept fixed~\cite{Braaten:2004rn}. In this limit, all the two-body observables scale with $a_{0}$. For $a_{0}>0$, there exist a shallow bound state with the radius $R=a_{0}$, which indicates $X=1$. Thus, the bound state in the scaling limit is always dominated by the composite structure~\cite{Hyodo:2014bda,Hanhart:2014ssa}. In the EFT description, $g_{0}\to 0$ corresponds to the scaling limit, where Eq.~\eqref{eq:X-bound} indicates $X=1$. The relation~\eqref{eq:comp-rel-bound} shows that the leading violation of the scaling $a_{0}=R$ is expressed by $X$.

We emphasize that the EFT used here is \textit{not} a model, but a universal description of the low-energy phenomena. The separable nature of the interaction, which was assumed in Ref.~\cite{Sekihara:2014kya}, is a consequence of the contact interaction of the EFT. 

\section{Model-independent relations for unstable quasi-bound states}
To generalize the result to the quasibound state, we introduce additional scattering channel $\psi^{\prime}\phi^{\prime}$ into which the bound state decays. We consider the Hamiltonian $H=H_{\mathrm{free}} + H_{\mathrm{free}}^{\prime} +H_{\mathrm{int}}+H_{\mathrm{int}}^{\prime}$ with
\begin{align}
H_{\mathrm{free}}^{\prime} 
&=\int d\bm{r}
\biggl[\frac{1}{2 M^{\prime}} \mathbf{\nabla} \psi^{\prime\dagger} \cdot\mathbf{\nabla} \psi^{\prime} 
- \nu_{\psi} \psi^{\prime\dagger} \psi^{\prime}
 \nonumber \\
&\quad 
+\frac{1}{2 m^{\prime}} \mathbf{\nabla} \phi^{\prime\dagger} \cdot\mathbf{\nabla} \phi^{\prime} -\nu_{\phi} \phi^{\prime\dagger} \phi^{\prime}
\biggr] , \\
H_{\mathrm{int}}^{\prime} &= \int d\bm{r}
\biggl[
g_{0}^{\prime} \left( B_0^\dagger \phi^{\prime}\psi^{\prime} + \psi^{\prime\dagger}\phi^{\prime\dagger} B_0 \right) + v_0^{\prime} \psi^{\prime\dagger}\phi^{\prime\dagger} \phi^{\prime}\psi^{\prime} \nonumber \\
&\quad +v_0^{t} 
(\psi^{\dagger}\phi^{\dagger} \phi^{\prime}\psi^{\prime}
+\psi^{\prime\dagger}\phi^{\prime\dagger} \phi^{}\psi^{}) 
\biggr] ,
\end{align}
with $\nu_{0}^{\prime}=\nu_{\psi} +\nu_{\phi}>0$. The eigenstate of the two-body problem is written as
\begin{align}
   \ket{\Psi} 
   &=c\ket{B_{0}}+\int\frac{d\bm{p}}{(2\pi)^{3}} 
   \bigl[\chi(\bm{p})\ket{\bm{p}}
   +\chi^{\prime}(\bm{p})\ket{\bm{p}^{\prime}}
   \bigr]
   \label{eq:twobody2} ,
\end{align}
where $\ket{\bm{p}^{\prime}}=\tilde{\psi}^{\prime\dag}(\bm{p})\tilde{\phi}^{\prime\dag}(-\bm{p})/\sqrt{\mathcal{V}}\ket{0}$ denotes the scattering state of the $\psi^{\prime}\phi^{\prime}$ system.

The compositeness of the unstable particles should be defined with caution~\cite{Hyodo:2013nka}. The quasi-bound state is the eigenstate of the Hamiltonian with a complex eigenvalue
\begin{align}
   H\ket{QB} &= E_{QB}\ket{QB},
   \quad E_{QB}\in \mathbb{C} .
\end{align}
The imaginary part of the eigenenergy represents the decay width of the quasibound state. The normalization should be given by introducing the Gamow vector $\ket{\overline{QB}}=\ket{QB}^{*}$ as $\bra{\overline{QB}}\kket{QB}=1$~\cite{Kukulin}. This leads to
\begin{align}
   Z &= \bra{B_{0}}\kket{QB}^{2},
   \quad 
   X^{(\prime)} = \int \frac{d\bm{p}}{(2\pi)^{3}} \bra{\bm{p}^{(\prime)}}\kket{QB}^{2}, \\
   Z&+X+X^{\prime} =1,\quad Z,X,X^{\prime} \in \mathbb{C} .
\end{align}
Thus, the compositeness and elementariness are in general complex. This is the property inherent in the unstable states. 

The forward scattering amplitude and the compositeness of the $\psi\phi$ channel are given by Eqs.~\eqref{eq:amp-bound} and \eqref{eq:X-bound} with replacing $-B$ by $E_{QB}$ and using
\begin{align}
   v(E) &= v_{0}+\frac{g_{0}^{2}}{E-\nu_{0}}
   +\frac{\left(v_{0}^{t}+\frac{g_{0}g_{0}^{\prime}}{E-\nu_{0}}\right)^{2}}
   {[\bar{G}(E)]^{-1}-(v_{0}^{\prime}+\frac{g_{0}^{\prime 2}}{E-\nu_{0}})} 
   \label{eq:vE} , \\
   \bar{G}(E)
   &= \frac{1}{2\pi^2} 
\int_{0}^{\Lambda}dp \frac{p^{2}}{E-p^2/(2\mu^{\prime})+\nu_{0}^{\prime}+i0^{+}} , 
\end{align}
where $\mu^{\prime}=M^{\prime}m^{\prime}/(M^{\prime}+m^{\prime})$. We define the ``radius'' of the quasi-bound state by the analytic continuation of that of the bound state
\begin{align}
   R &= 1/\sqrt{-2\mu E_{QB}} \in \mathbb{C} .
\end{align}
Expanding the scattering length of the $\psi\phi$ channel $a_{0}$ in powers of $1/|R|$, we obtain the following relation
\begin{align}
a_0
&=  R \Biggl\{\frac{2X}{1+X} + {\mathcal O}\left(\left|\tfrac{R_{\mathrm{typ}}}{R}\right| \right) + \sqrt{\frac{\mu^{\prime 3}}{\mu^{3}}} \mathcal{O} \left( \left| \tfrac{l}{R} \right|^{3}\right) \Biggr\}
\label{eq:comp-rel-quasi} .
\end{align}

The first two terms are obtained from the expansion of $G(E)$ in the same way with the bound state case. The last term comes from the expansion of the additional contribution $\bar{G}(E)$ in $[v(E)]^{-1}$. Because the effective range expansion is assumed to be valid at $E=E_{QB}$, the expansion of $\bar{G}(E)$ should start from the linear term in $E_{QB}$, namely, $ \mathcal{O} (R^{-2})$, which involves the new length scale $l\equiv 1/\sqrt{2\mu\nu_{0}^{\prime}}$ associated with the energy difference of the thresholds. The relation for the effective range is obtained similarly. We emphasize that Eq.~\eqref{eq:comp-rel-quasi} for the unstable states cannot be obtained in the derivation of Ref.~\cite{Weinberg:1965zz} which uses the eigenstate expansion of the full Hamiltonian.

In this way, we find that the leading term for the quasi-bound state with large $|R|$ has the same form with the bound state relation, but for complex-valued $a_{0}$, $R$, and $X$. The $\mathcal{O} ( \left| l/R \right|^{3})$ term can be neglected if the threshold energy of the decaying channel is sufficiently apart from the two-body channel of interest. The same argument can be applied to the case with $\re E_{QB}>0$, as long as $|R|$ is large. We note that the model-independent relation determines only $X$ and $1-X=Z+X^{\prime}$. It is not possible to separate $Z$ and $X^{\prime}$ in a model-independent manner. In the following, we replace $Z+X^{\prime}\to Z$ and discuss the interpretation of complex $Z$ and $X$ with $Z+X=1$. 

\section{Interpretation of complex $Z$ and $X$}
The interpretation of the complex norm of unstable particles is a long-standing problem. In general, complex $Z$ and $X$ cannot be interpreted as probabilities, because the values are not bounded. Strictly speaking, only the real and bounded quantities can be regarded as probabilities. As pointed out in Ref.~\cite{Sekihara:2014kya}, however, reasonable  interpretation can be given if $Z$ and $X$ have similarity with those of the bound states, i.e., $|\im Z|,|\im X|\ll 1$ and $0\lesssim\re Z, \re X \lesssim 1$. In other words, if the violation of the condition~\eqref{eq:ZXbound} is small, we can make a reasonable interpretation. For a quantitative discussion, we note the triangular inequality:
\begin{align}
    |Z|+|X| \geq &
    |Z+X|=1 ,
\end{align}
where the equality holds if and only if Eq.~\eqref{eq:ZXbound} is satisfied. We thus define 
\begin{align}
    U \equiv |Z| +|X| -1  \label{eq:kaisyaku2-U}
\end{align}
which quantifies the deviation from the bound state limit, Eq.~\eqref{eq:ZXbound}. Because the solid interpretation is possible for the bound state, the quantity $U$ is understood as the uncertainty of the interpretation of the complex $X$. A similar discussion can be found in Ref.~\cite{PL33B.547}. 

Next, we define real and bounded quantities $\tilde{X}$ and $\tilde{Z}$ from complex $X$ and $Z$. Our aim is to interpret $\tilde{X}$ as the probability of finding the composite component, when $U$ is small. For the probabilistic interpretation, $\tilde{X}$ and $\tilde{Z}$ should satisfy
\begin{align}
\tilde{Z}+\tilde{X} =1,\quad \tilde{Z},\tilde{X} \in [0,1] .
\label{eq:ZXQB}
\end{align}
In addition, $\tilde{X}$ ($\tilde{Z}$) should reduce to $X$ ($Z$) in the bound state limit, which has a clear meaning of the probability. All these conditions can be satisfied by defining
\begin{align}
    \tilde{Z} \equiv &\frac{1 - |X| + |Z|}{2},
\quad \tilde{X} \equiv \frac{1 - |Z| + |X|}{2}, \label{eq:kaisyaku2-X}
\end{align} 
Hence, we regard $\tilde{Z}$ ($\tilde{X}$) as the elementariness (compositeness), provided that the uncertainty $U$ is small.\footnote{Reference~\cite{PL33B.547} (Ref.~\cite{Aceti:2014ala}) suggests to consider $|X|-U=1-|Z|$ ($\re X$) as the probability of compositeness, but this quantity can be negative. In contrast, our definition of $\tilde{X}$ is positive definite.} 

In this way, we establish the framework to quantitatively interpret the structure of unstable particles. However, we may in general encounter the case with relatively large $U$, because there is no restriction for the values of the complex $X$ and $Z$. If the uncertainty $U$ is not small, we need to adopt alternative method to clarify the structure. A useful quantity is the ratio $|r_{e}/a_{0}|$. As discussed in Ref.~\cite{Hyodo:2013iga}, when the elementary component is large, the effective range $r_{e}$ increases. The ratio behaves as $|r_{e}/a_{0}|=0,\, 1.5, \,\infty$ for $X=1,\, 0.5,\, 0$, respectively. Thus, the ratio should be small when the quasibound state is dominated by the composite structure. This criterion is essentially the same with the pole counting argument~\cite{Morgan:1990ct,Morgan:1992ge}, as discussed in Refs.~\cite{Baru:2003qq,Hyodo:2013iga,Hyodo:2014bda}. We however note that this type of analysis gives only a qualitative statement.

\section{Other approaches for compositeness}
There have been several studies on the compositeness of hadron resonances~\cite{Baru:2003qq,Baru:2010ww,Hyodo:2011qc,Aceti:2012dd,Hyodo:2013iga,Chen:2013upa,Hyodo:2013nka,Sekihara:2014kya,Guo:2015daa}. In the pioneering study of Refs.~\cite{Baru:2003qq,Baru:2010ww}, the compositeness is expressed by the integration of the spectral density. In more recent works~\cite{Hyodo:2011qc,Aceti:2012dd,Hyodo:2013nka,Sekihara:2014kya,Guo:2015daa}, the compositeness is given by the product of the residue of the amplitude and the derivative of loop function, which is essentially equivalent to Eq.~\eqref{eq:X-bound}. In the case of bound states, the weak-binding formula~\eqref{eq:comp-rel-bound} is derived from this expression in Ref.~\cite{Sekihara:2014kya}. The use of EFT in the context of the compositeness can be found in Ref.~\cite{Chen:2013upa}.

Our main result~\eqref{eq:comp-rel-quasi} is a direct generalization of the formula~\eqref{eq:comp-rel-bound} of Ref.~\cite{Weinberg:1965zz}. We show that the EFT description can be used not only for the description of the scattering amplitude, but also for the definition of the compositeness. In addition, the decay channels are explicitly introduced in EFT to describe the unstable eigenstates. In this way, the compositeness of the unstable state is expressed in terms of the observable quantities ($a_{0}$ and $E_{QB}$) in the weak binding limit. 

\section{Applications to exotic hadrons}
We have shown that the compositeness $X$ of the quasi-bound state can be model-independently evaluated by Eq.~\eqref{eq:comp-rel-quasi} when the correction terms are small and the eigenenergy $E_{QB}$ and the scattering length $a_{0}$ are given. Now we apply this framework for the structure of near-threshold exotic hadrons. We utilize empirical determinations of $E_{QB}$ and $a_{0}$ by several existing data analyses. For a given set of $E_{QB}$ and $a_{0}$, we estimate the correction terms and the uncertainty of the interpretation $U$. Different input values of $E_{QB}$ and $a_{0}$ induce the systematic uncertainty of the results, which is rooted in the precision of the empirical determination.

The $\Lambda(1405)$ resonance is a negative parity excited baryon 
which lies close to the $\bar{K}N$ threshold and decays into the $\pi\Sigma$ channel. The threshold parameters of $\Lambda(1405)$ have recently been determined by the detailed studies of the experimental data around the $\bar{K}N$ threshold with the chiral effective theories~\cite{Ikeda:2011pi,Ikeda:2012au,Mai:2012dt,Guo:2012vv,Mai:2014xna} in which the eigenenergies are found in the region $|R|\gtrsim 1.5$ fm. The correction terms are found to be small, $|R_{\rm typ}/R|\lesssim 0.17$ and $|l/R|^{3}\lesssim 0.14$, where the $\bar{K}N$ interaction range is estimated by the $\rho$ meson exchange. From the central values of $E_{QB}$ and $a_{0}$ in these analyses, we determine the $\bar{K}N$ compositeness as summarized in Table~\ref{tab:Lambda}.\footnote{The scattering lengths of Refs.~\cite{Ikeda:2012au,Guo:2012vv} are obtained from the isospin averaged amplitude. Others are evaluated at the $K^{-}p$ threshold by $a_{0}=(a_{K^{-}p}+a_{\bar{K}^{0}n})/2$.} We find that $\tilde{X}$ is close to unity in all cases, indicating the $\bar{K}N$ composite structure of $\Lambda(1405)$. Although some results are associated with $U\sim 0.6$, the ratio $|r_{e}/a_{0}|<1.5$ is consistent with the $\bar{K}N$ composite dominance. 

\begin{table}[bt]
	\begin{center}
		\begin{ruledtabular}
		\begin{tabular}{llllclr}
             Ref. & $E_{QB}\physdim{MeV}$ & $a_0 \physdim{fm} $ & $X_{\bar{K}N}$ & $\tilde{X}_{\bar{K}N}$ & $U$ & $|r_{e}/a_{0}|$  \\  \hline
			 \cite{Ikeda:2012au}  & $-10-i26$ & $1.39 - i 0.85$ 
			 & $1.2+i0.1$ & $1.0$ & $0.5$ & $0.2$  \\ 
			 \cite{Mai:2012dt}  & $-\phantom{0}4-i\phantom{0}8$ & $1.81-i0.92$ 
			 & $0.6+i0.1$ & $0.6$ & $0.0$ & $0.7$ \\ 
			 \cite{Guo:2012vv}  & $-13-i20$ & $1.30-i0.85$ 
			 & $0.9-i0.2$ & $0.9$ & $0.1$ & $0.2$  \\
			 \cite{Mai:2014xna}  & $\phantom{-0}2-i10$ & $1.21-i1.47$ 
			 & $0.6+i0.0$ & $0.6$ & $0.0$ & $0.7$ \\ 
			 \cite{Mai:2014xna}  & $-\phantom{0}3-i12$ & $1.52-i1.85$ 
			 & $1.0+i0.5$ & $0.8$ & $0.6$ & $0.4$ \\ 
		\end{tabular} 
		\caption{Properties and results for $\Lambda (1405)$. Shown are the eigenenergy $E_{QB}$, $\bar{K}N(I=0)$ scattering length $a_{0}$, the $\bar{K}N$ compositeness $X_{\bar{K}N}$ and $\tilde{X}_{\bar{K}N}$, uncertainty of the interpretation $U$, and the ratio of the effective range to the scattering length $|r_{e}/a_{0}|$. The scattering length is defined as $a_{0}=-f(E=0)$.}
		\label{tab:Lambda}
        \end{ruledtabular}
	\end{center}
\end{table}

Near the $\bar{K}K$ threshold, there are two scalar mesons, $a_{0}(980)$ and $f_{0}(980)$ with the isospin $I=1$ and $I=0$, respectively. The decay channel of $a_{0}(980)$ [$f_{0}(980)$] is $\pi\eta$ ($\pi\pi$). As summarized in Ref.~\cite{Sekihara:2014qxa}, recent experimental data around the $\bar{K}K$ threshold has been analyzed by Flatte parametrization~\cite{Adams:2011sq,Ambrosino:2009py,Bugg:2008ig,Achasov:2000ku,Teige:1996fi,Aaltonen:2011nk,Ambrosino:2005wk,Garmash:2005rv,Ablikim:2004wn,Link:2004wx,Achasov:2000ym}, from which $E_{QB}$ and $a_{0}$ can be determined. Except for Ref.~\cite{Adams:2011sq}, the obtained eigenenergies satisfy $|R|\gtrsim 1.5$ fm. Estimating $R_{\rm typ}$ by the $\rho$ exchange, we find $|R_{\rm typ}/R|\lesssim 0.17$ and $|l/R|^{3}\lesssim 0.04$ for both mesons (with Ref.~\cite{Adams:2011sq}, we obtain $|R_{\rm typ}/R|\sim 0.25$ and  $|l/R|^{3}\sim 0.13$). The evaluated $\bar{K}K$ compositeness are summarized in Table~\ref{tab:a0} (Table~\ref{tab:f0}) for $a_{0}(980)$  [$f_{0}(980)$], where we find that the uncertainty $U$ is small for all cases. The results of $a_{0}(980)$ show that $\tilde{X}$ is small and $|r_{e}/a_{0}|$ is much larger than $1.5$, except for Ref.~\cite{Achasov:2000ku}. Given the large uncertainty of the parameters in Ref.~\cite{Achasov:2000ku} (see Ref.~\cite{Sekihara:2014qxa}), we conclude that the structure of $a_{0}(980)$ is dominated by the non-$\bar{K}K$ component. On the other hand, the results of $f_{0}(980)$ are scattered and not conclusive, as a consequence of the  uncertainties of the Flatte parameters. We emphasize that the true values of the Flatte parameters must be unique. The large deviation of the results in Table~\ref{tab:f0} originates in the large uncertainty of the determination of the Flatte parameters in Refs.~\cite{Aaltonen:2011nk,Ambrosino:2005wk,Garmash:2005rv,Ablikim:2004wn,Link:2004wx,Achasov:2000ym}. The small values of $U$ in Table~\ref{tab:f0} indicates that, if the values of the Flatte parameters are determined unambiguously, then the complex compositeness of $f_0(980)$ can be interpreted with very little uncertainty.

\begin{table}[bt]
	\begin{center}
		\begin{ruledtabular}
		\begin{tabular}{llllclr}
             Ref. & $E_{QB} \physdim{MeV}$ & $a_0 \physdim{fm} $ & $X_{\bar{K}K}$ & $\tilde{X}_{\bar{K}K}$ & $U$ & $|r_{e}/a_{0}|$  \\  \hline
			 \cite{Adams:2011sq}  & $31-i70$ & $-0.03 - i 0.53$ 
			 & $0.2-i0.2$ & $0.3$ & $0.1$ & $4.8$  \\ 
			 \cite{Ambrosino:2009py} & $\phantom{0}3-i25$ & $\phantom{-}0.17 - i 0.77$ 
			 & $0.2-i0.2$ & $0.2$ & $0.1$ & $6.5$  \\ 
			 \cite{Bugg:2008ig} & $\phantom{0}9-i36$ & $\phantom{-}0.05 - i 0.63$ 
			 & $0.2-i0.2$ & $0.2$ & $0.1$ & $7.2$  \\ 
			 \cite{Achasov:2000ku} & $14-i\phantom{0}5$ & $-0.13 - i 2.19$ 
			 & $0.8-i0.4$ & $0.7$ & $0.3$ & $0.5$  \\ 
			 \cite{Teige:1996fi} & $15-i29$ & $-0.13 - i 0.52$ 
			 & $0.1-i0.2$ & $0.1$ & $0.1$ & $13\phantom{.0}$  \\ 
		\end{tabular} 
		\caption{Properties and results for $a_{0} (980)$. Shown are the eigenenergy $E_{QB}$, $\bar{K}K(I=1)$ scattering length $a_{0}$, the $\bar{K}K$ compositeness $X_{\bar{K}K}$ and $\tilde{X}_{\bar{K}K}$, uncertainty of the interpretation $U$, and the ratio of the effective range to the scattering length $|r_{e}/a_{0}|$.}
		\label{tab:a0}
        \end{ruledtabular}
	\end{center}
\end{table}

\begin{table}[t]
	\begin{center}
		\begin{ruledtabular}
		\begin{tabular}{llllclr}
             Ref. & $E_{QB} \physdim{MeV}$ & $a_0 \physdim{fm} $ & $X_{\bar{K}K}$ & $\tilde{X}_{\bar{K}K}$ & $U$ & $|r_{e}/a_{0}|$  \\  \hline
			 \cite{Aaltonen:2011nk}  & $\phantom{-}19-i30$ & $0.02 - i 0.95$ 
			 & $0.3-i0.3$ & $0.4$ & $0.2$ & $2.6$  \\ 
			 \cite{Ambrosino:2005wk}  & $-\phantom{0}6-i10$ & $0.84 - i 0.85$ 
			 & $0.3-i0.1$ & $0.3$ & $0.0$ & $5.4$  \\ 
			 \cite{Garmash:2005rv}  & $-\phantom{0}8-i28$ & $0.64 - i 0.83$ 
			 & $0.4-i0.2$ & $0.4$ & $0.1$ & $2.1$  \\ 
			 \cite{Ablikim:2004wn}  & $\phantom{-}10-i18$ & $0.51 - i 1.58$ 
			 & $0.7-i0.3$ & $0.6$ & $0.1$ & $0.7$  \\ 
			 \cite{Link:2004wx}  & $-10-i29$ & $0.49 - i 0.67$ 
			 & $0.3-i0.1$ & $0.3$ & $0.0$ & $4.0$  \\ 
			 \cite{Achasov:2000ym}  & $\phantom{-}10-i\phantom{0}7$ & $0.52 - i 2.41$ 
			 & $0.9-i0.2$ & $0.9$ & $0.1$ & $0.2$  \\ 
		\end{tabular} 
		\caption{Properties and results for $f_{0} (980)$. Shown are the eigenenergy $E_{QB}$, $\bar{K}K(I=0)$ scattering length $a_{0}$, the $\bar{K}K$ compositeness $X_{\bar{K}K}$ and $\tilde{X}_{\bar{K}K}$, uncertainty of the interpretation $U$, and the ratio of the effective range to the scattering length $|r_{e}/a_{0}|$.}
		\label{tab:f0}
        \end{ruledtabular}
	\end{center}
\end{table}

The $\bar{K}N$ compositeness of $\Lambda(1405)$ and the $\bar{K}K$ compositeness of the scalar mesons have been evaluated in Refs.~\cite{Sekihara:2014kya,Sekihara:2014qxa} with explicit model calculations, which are in good agreement with the present model-independent results. The $\bar{K}N$ composite dominance of $\Lambda(1405)$ is supported by the recent lattice QCD calculation~\cite{Hall:2014uca} and the realistic $\bar{K}N$ potential~\cite{Miyahara:2015bya}. The derivation of the model-independent relation~\eqref{eq:comp-rel-quasi} enables us to draw this conclusion only from the experimentally observable quantities, the eigenenergy and the scattering length.

We have demonstrated that the generalization of the model-independent weak-binding relation to quasi-bound states is a powerful tool for unveiling the structure of exotic hadrons. Once the scattering length and the eigenenergy are determined, the same method can be applied to the exotic hadrons in the heavy sector~\cite{Swanson:2006st,Brambilla:2010cs}. To this end, it is important to determine the threshold parameters, for instance, by the Flatte parametrization or by the lattice QCD technique. We also note that the  isospin breaking effect may become important near the threshold. The generalization of the present work for this case is in progress~\cite{Kamiya2}.

\section{Acknowledgments}
The authors thank Jose Antonio Oller for the valuable discussion of the compositeness and the details of Ref.~\cite{Guo:2012vv}, Takayasu Sekihara for useful information of the scalar mesons and Maxim Mai for providing the scattering length of the analysis in Ref.~\cite{Mai:2014xna}. This work is supported in part by JSPS KAKENHI Grant No. 24740152 and by the Yukawa International Program for Quark-Hadron Sciences (YIPQS).


%

\end{document}